\documentclass{article}
\usepackage{authblk}

\usepackage{natbib}
\usepackage[margin=1in]{geometry}
\usepackage{enumitem}
\usepackage{amsmath,amsfonts,amssymb,amsthm}
\usepackage{multicol}
\usepackage{multirow}
\usepackage{setspace}
\usepackage{caption}

\usepackage{hyperref}
\hypersetup{colorlinks=false}

\usepackage{todo}
\usepackage{tikz}
\usetikzlibrary{shapes,arrows}

\usepackage{titlesec}
\titleformat{\section}{\large\bfseries}{\thesection}{1em}{}
\titleformat{\subsection}{\normalsize\bfseries}{\thesection}{1em}{}

\usepackage{everypage}

\usepackage{listings}
\usepackage{color}
\usepackage{textcomp}
\definecolor{listinggray}{gray}{0.9}
\definecolor{lbcolor}{rgb}{1,1,1}
\newcommand{\kma}{\texttt{kma} }

\begin{document}
\author[1]{Harold Pimentel}
\author[2]{John G. Conboy}
\author[3]{Lior Pachter\thanks{lpachter@math.berkeley.edu}}
\affil[1]{Department of Computer Science, UC Berkeley.}
\affil[2]{Biological Systems and Engineering Division, Lawrence Berkeley National Laboratory.}
\affil[3]{Departments of Computer Science, Mathematics and Molecular \& Cell Biology, UC Berkeley.}

\title{Keep Me Around: Intron Retention Detection and Analysis}




\maketitle

\begin{abstract}

\textbf{Summary:} We present a tool, \textbf{k}eep \textbf{m}e \textbf{a}round
(\texttt{kma}), a suite of python scripts and an R package that finds retained introns
in RNA-Seq experiments and incorporates biological replicates to reduce the
number of false positives when detecting retention events. \kma uses the
results of existing quantification tools that probabilistically assign
multi-mapping reads, thus interfacing easily with transcript quantification
pipelines. The data is represented in a convenient, database style format that
allows for easy aggregation across introns, genes, samples, and conditions to
allow for further exploratory analysis.

\textbf{Availability:} The source code is available under the GPLv2 license and can be found at:
\newline \href{http://github.com/pachterlab/kma}{http://github.com/pachterlab/kma}

\textbf{Contact:} \href{lpachter@math.berkeley.edu}{lpachter@math.berkeley.edu}
\end{abstract}


\section{Motivation}

Many organisms exhibit intron retention events that can be measured with
RNA-Seq \citep{burgess_alternative_2014,braunschweig_widespread_2014}, and
recent publications suggest that these events are important constituents of
transcriptome regulation. While some existing tools can detect intron retention
events \citep{miso, dexseq, bai_ircall_2015}, none that we are aware of
incorporate biological replicates to reduce the reporting of false-positives.
Other tools have been mentioned in the literature, but do not have freely
available software \citep{nascentSeq, rasko2013, braunschweig_widespread_2014, sharp}.
There is therefore a need for a robust intron retention detection method that
is based on rigorous quantification of intron retention followed by assessment
of significance using biological replicates.

We present \textbf{k}eep \textbf{m}e \textbf{a}round (\texttt{kma}), a set of tools for detecting intron retention in
RNA-Seq experiments that utilizes biological replicates to improve accuracy.
\kma currently uses the transcript quantification method eXpress
\citep{eXpress}, but is compatible with with any RNA-Seq quantification pipeline.


\section{Implementation}

\begin{figure*}[!tp]

  \centerline{\includegraphics[width=\linewidth]{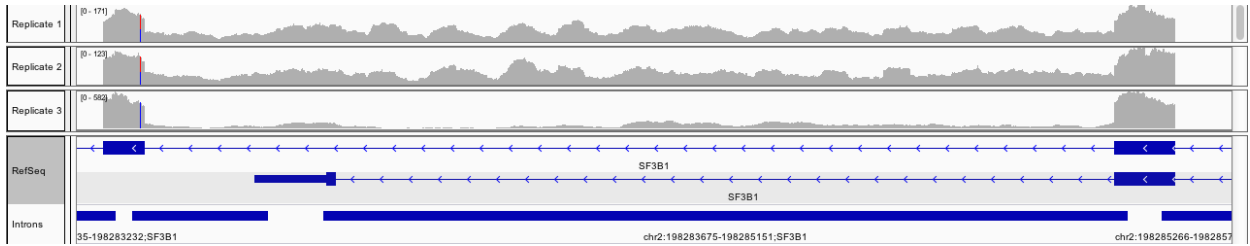}}

  \caption{Example of intron retention in orthochromatic erythroblasts from
    \citep{pimentel2014} in gene SF3B1. The ``Intron" track shows the regions of
    included pseudo-transcripts and that each pseudo-transcript overlaps neighboring exons by 25 bases. The coverage track shows that the first few
    introns are covered by very few reads, but intron 4 (chr2:198283675-198185151) shows intron retention in all
    replicates.}\label{fig:measurable}

\end{figure*}

\kma begins by performing a pre-processing step consisting of several python
scripts that find ``measurable" intronic regions called inclusion regions
(regions in which none of the overlapping isoforms contain an exon), together
with the corresponding isoforms which could retain the intron called overlap
isoforms. \kma then outputs a table of intron-transcript relationships. This
table includes the (1) intron coordinates, (2) intron quantification
coordinates (3) transcripts which could potentially retain the intron, and (4)
the gene name. The intron quantification coordinates differ from the exact
intron coordinates by including a small region of the neighboring exons which
is several bases shorter than the read length (Figure \ref{fig:measurable}).
This exonic overlap ensures that the reads spanning the intron-exon junctions
are included into the intron expression. These reads are often valuable
information; if they are unique, they give strong evidence for the expression
of the intron. \kma also outputs a BED track containing intronic quantification
coordinates, as well as a FASTA file containing the intronic sequences to
quantify against. This pre-processing step only has to be performed once
assuming the transcriptome annotation does not change and read size is at least
a few bases longer than exon overlap.

\kma is designed to leverage existing transcript quantification methods. This
allows for the computation of relative abundance of introns as well as
transcripts while allowing multi-mapping reads to be processed using well
understood models already developed in existing tools \citep{PachterArXiv, rsem2010}.
After the pre-processing step, the intronic sequences are added to the
transcriptome and the chosen quantification method is run using the augmented
transcriptome. Any method can be used, provided it outputs expression in a unit
that is additive, e.g. transcripts per million (TPM).

Once introns and transcripts are quantified from all samples in the experiment,
the data can be post-processed and further analyzed in an {R} package
\citep{rstats} that is part of {kma}.  We currently provide functions to read
data from eXpress, but it is quite simple to add a new function that reads in
other formats; all that is required is the target identifier and corresponding
expression estimate. Once data is read in, retention is computed by taking the
intron expression (numerator) and summing the expression of the overlapping
transcripts plus the intron expression (denominator). This calculation leads to
a natural measurement of intron retention, the proportion of the transcript
expression containing the intron, also known as the proportion spliced in
\citep{miso}.

While we store a special object of class {IntronRetention}, the majority
of the operations depend only on the data stored in database-like data frames
with each row being an intron observation from one sample.  A common row
contains categorical fields {intron, sample, condition} which serve as
a key, along with measurements {retention, numerator, denominator,
  unique\_reads} and various columns for filters. This allows for fast
aggregation and manipulation via packages such as dplyr \citep{dplyr}.
Summaries of retention across subgroups such as specific introns, conditions,
or samples can be quickly computed by simple queries. We provide
common summaries as functions, but the raw data frame is always
available for further analysis. In addition to easy manipulation, this data
format is suitable for exploratory analysis in plotting tools such as
ggplot2 \citep{ggplot2}.

In certain situations, estimates of the retention level can be unreliable due
to low coverage in the exonic regions, or high variations in coverage due to
biases or repetitive sequences. Coverage filters were implemented based on
relative expression or rank, along with a ``zero coverage filter''. The zero
coverage filter finds the longest spanning region in an intron which has no
reads starting in that region. Then, it computes the probability of observing
a region of length $Z$ with no reads starting in it, given the intron's
expression. The intron is removed from consideration if the probability is low.


Unlike other publicly available intron retention tools which simply provide an estimate for intron
retention per sample, our method provides a resampling hypothesis testing
procedure to determine whether the mean is greater than what one would expect due to
reshuffling of the given data in those samples. The null distribution is
generated from the filtered list by randomly selecting a retention value from
each sample per condition $B$ times. For each set of samples, the mean is
computed. After the null distribution is generated, the $p$-value is computed
by finding the proportion of null values that the observed mean is greater
than. This allows for a lower false-positive rate when detecting IR events.
This procedure also helps shield against samples that have contamination of non-mature mRNA.


\section{Discussion}

We have developed an R package \kma that addresses the issue of finding
intron retention events. Since this tool only slightly modifies existing
quantification pipelines by introducing an augmented transcriptome, \kma can
easily be introduced into existing RNA-Seq quantification pipelines. In R, the
data is represented in a database-like format that allows for flexible and fast
aggregation allowing for exploratory analysis to be carried out relatively
easily. We also implemented a hypothesis testing procedure which reduces the
likelihood of finding false-positive intron retention events by incorporating
biological replicates.


\section*{Acknowledgement}
HP was supported by the NSF GRFP. JGC and LP were supported in part by NIH R01 DK094699.

\bibliographystyle{plain}


\bibliography{document}

\end{document}